\documentclass{article}

\usepackage{graphicx} 
\usepackage[utf8]{inputenc}
\usepackage{amsmath, amssymb, amsfonts}
\usepackage{bbm}
\usepackage{float}
\usepackage{booktabs}
\usepackage{multirow}
\usepackage{geometry}
\usepackage{xcolor}
\usepackage{microtype}
\usepackage{hyperref}

\usepackage[style=numeric-comp, sorting=none]{biblatex} 
\title{Quark masses and mixing from Modular $S'_4$ with Canonical K\"ahler Effects}

\author{
	Ivo~de~Medeiros~Varzielas$^{1}$\thanks{E-mail: ivo.de@udo.edu}  , 
    Manuel~Paiva$^{2}$\thanks{E-mail: up202002826@up.pt}
	\\
	{\small $^1$ CFTP, 
		Instituto Superior T\'{e}cnico, Universidade de Lisboa, Av. Rovisco Pais 1, 1049-001 Lisboa, Portugal}\\
    {\small $^2$ DFA, Faculdade de Ciências da Universidade do Porto, Rua do Campo Alegre, s/n, 4169-007 Porto, Portugal
		}
    }

\date{April 2026}

\begin{document}

\maketitle
\begin{abstract}
    We propose a quark flavor model based on modular $S'_4$ with a general CP symmetry. 
    CP violation in the quark sector is entirely realized by the modulus $\langle \tau \rangle$. We show that the canonical normalization induced by the K\"ahler metric plays a crucial role in reproducing the observed hierarchies, while maintaining coupling constants of order $\mathcal{O}(1)$. The minimal model achieves a great fit to the quark sector data, which we take as the PDG 2024 data extrapolated to the GUT scale.
\end{abstract}

\section{Introduction}

In the Standard Model (SM), fermion masses and mixing angles arise from Yukawa couplings, which are fundamentally free parameters that must be fitted to experimental data. The absence of an underlying theoretical principle to dictate these values constitutes the longstanding Flavor Problem. Specifically, this leaves several critical questions unanswered. Why do charged fermions exhibit such extreme mass hierarchies, with the top quark being roughly five orders of magnitude heavier than the electron? Why is the quark mixing pattern small and hierarchical, while the lepton mixing features  large angles? Furthermore, the observation of neutrino oscillations definitively proves that neutrinos possess tiny but non-zero masses, a feature not predicted by the original SM formulation.

In recent years, modular flavor symmetries \cite{feruglio2017neutrinomassesmodularforms,kobayashi2023modularflavorsymmetricmodels} have emerged as an attractive solution to address this problem, by generating these mixing patterns from the vacuum expectation value of a modulus field. A common limitation of many flavor models is the requirement of non-natural Yukawa-type coupling constants, that deviate significantly from $\mathcal{O}(1)$ values \cite{okada2020quarkleptonflavorscommon,arriagaosante2024quarkleptonmodularmodels,Qu_2025,petcov2026sprime4quarkflavourmodel}. The necessity of using artificial hierarchies between parameters reintroduces an issue flavor models were trying to solve, rendering them at best as a partial solution to the flavor problem.

Recent experimental progress has significantly improved the precision of the flavor observables, requiring models to be as precise as possible.
The tight margins of the Particle Data Group (PDG) 2024 data \cite{ParticleDataGroup:2024cfk}, even when extrapolated to the GUT scale \cite{antusch2026updatedrunningquarklepton}, make it significantly more challenging to construct successful quark flavor models.
The improvement is illustrated in table \ref{tab:2024vs2022}.

\begin{table}[H]
\centering
\renewcommand{\arraystretch}{1.5}
\resizebox{\textwidth}{!}{
\begin{tabular}{|c|c|c|c|}
\hline
\textbf{SM Quantity} & \textbf{2022 PDG (Value $\pm$ Error)}  & \textbf{2024 PDG (Value $\pm$ Error)}  & \textbf{Error Width Reduction} \\ \hline
$y_u/10^{-6}$ & $2.79^{+0.60}_{-0.36}$ & $2.76 \pm 0.07$ & $\sim 85\%$ (6.8x more precise) \\ \hline
$y_d/10^{-5}$ & $0.49^{+0.05}_{-0.02}$ & $0.49 \pm 0.01$ & $\sim 71\%$ (3.5x more precise) \\ \hline
$y_s/10^{-4}$ & $0.99^{+0.08}_{-0.05}$ & $0.98 \pm 0.01$ & $\sim 85\%$ (6.5x more precise) \\ \hline
$y_c/10^{-3}$ & $1.39^{+0.05}_{-0.04}$ & $1.39 \pm 0.03$ & $\sim 33\%$ (1.5x more precise) \\ \hline
$y_b/10^{-2}$ & $0.551^{+0.008}_{-0.006}$ & $0.550 \pm 0.006$ & $\sim 14\%$ reduction \\ \hline
$y_t$ & $0.4965^{+0.0078}_{-0.0081}$ & $0.4947 \pm 0.0079$ & No significant change \\ \hline
$\theta_{12}$ & $0.22702^{+0.00083}_{-0.00081}$ & $0.22702 \pm 0.00082$ & No significant change \\ \hline
$\theta_{23}/10^{-2}$ & $3.869^{+0.036}_{-0.041}$ & $3.874 \pm 0.038$ & No significant change \\ \hline
$\theta_{13}/10^{-3}$ & $3.41^{+0.08}_{-0.07}$ & $3.41 \pm 0.07$ & $\sim 6\%$ reduction \\ \hline
$\delta$ & $1.139 \pm 0.023$ & $1.139 \pm 0.023$ & No change \\ \hline
$J/10^{-5}$ & $2.65 \pm 0.07$ & $2.65 \pm 0.07$ & No change \\ \hline
\end{tabular}
}
\caption{Comparison of GUT-scale values ($M_\textrm{GUT}=2\times 10^{16}$ GeV) with $M_\textrm{SUSY} = 3$ TeV and $\tan\beta = 10$, emphasizing the reduction in uncertainties between the 2022 and 2024 PDG datasets \cite{antusch2026updatedrunningquarklepton}.}
\label{tab:2024vs2022}
\end{table}

Indeed, through an independent numerical analysis we confirmed that specific model predictions previously consistent with the 2022 dataset \cite{varzielas2023quarksmodulars4cusp} now exhibit tensions exceeding 4 standard deviations.
These highly restrictive bounds act as a stringent sieve, effectively falsifying a vast number of previously viable flavor models that can no longer accommodate the tight limits.

Here, we propose a flavor model based on the double cover of the $S_4$ \cite{Novichkov_2021} modular group, with a general CP (gCP) symmetry, and a minimal set of free parameters (real $\mathcal{O}(1)$ Yukawa-type constants). Our model is able to reproduce the mass hierarchies and mixing pattern of the quark  sector.
The CP symmetry is broken spontaneously solely by the complex vacuum expectation value (VEV) of the modulus field, $\tau$. Furthermore, the modular K\"ahler metric plays an important role in accounting for the mass hierarchies, through the distinct modular weights to the matter superfields.

The remainder of this paper is organized as follows. In Section \ref{sec:framework}, we introduce the mathematical framework of the modular symmetries and  the general form of the supersymmetric theory. In Section \ref{sec:model}, we present the model in terms of the choices of representations and weights, and the corresponding superpotential and mass matrices.
In Section \ref{sec:numerical-analysis}, we present the results from the best-fit parameters.

\section{Framework}\label{sec:framework}
\subsection{Modular symmetry}
The modular group $\Gamma\cong SL(2,\mathbb{Z}) $ is the special linear group
of $2 \times2$ integer matrices with unit determinant,

\begin{equation}
    \Gamma \equiv \mathrm{SL}(2, \mathbb{Z}) \equiv \left\{ \begin{pmatrix} a & b \\ c & d \end{pmatrix} \;\middle|\; a, b, c, d \in \mathbb{Z}, ad - bc = 1 \right\}.
\end{equation}

Although $\Gamma $ contains infinitely many elements, every element can be generated by three matrices

\begin{equation}
    S = \begin{pmatrix} 0 & 1 \\ -1 & 0 \end{pmatrix}, \quad 
T = \begin{pmatrix} 1 & 1 \\ 0 & 1 \end{pmatrix}, \quad 
R = \begin{pmatrix} -1 & 0 \\ 0 & -1 \end{pmatrix}.
\end{equation}

These generators satisfy the relations

\begin{equation}
\Gamma \cong \langle S, T, R \mid S^{2}=R,\; (ST)^{3}=\mathbbm{1},\; R^2=\mathbbm{1},\; RT=TR\rangle.
\end{equation}

$\Gamma(N)$ is called the Principal Congruence Subgroup of level N
\begin{equation}
    \Gamma(N) = \left\{ 
\begin{pmatrix} a & b \\ c & d \end{pmatrix} \in SL(2, \mathbb{Z}) : 
\begin{pmatrix} a & b \\ c & d \end{pmatrix} \equiv 
\begin{pmatrix} 1 & 0 \\ 0 & 1 \end{pmatrix} \pmod N 
\right\}.
\end{equation}

Although each $\Gamma(N)$ is infinite and discrete, these quotients
\begin{equation}
    \Gamma'_N=\Gamma/\Gamma(N),
\end{equation}
\begin{equation}
    \Gamma_N = \Gamma'_N/\{I,-I\},
\end{equation}
are finite and  known as  \emph{homogeneous finite modular group} and \emph{inhomogeneous finite modular group}, respectively.
\begin{equation}
\Gamma'_N \cong \langle S, T \mid S^4 = \mathbbm{1}, (ST)^3 = \mathbbm{1}, S^2 T = T S^2, T^N = \mathbbm{1} \rangle.
\end{equation}
\begin{equation}
\Gamma_N \cong \langle S, T \mid S^{2}=\mathbbm{1}, (ST)^{3}=\mathbbm{1}, T^N=\mathbbm{1}\rangle.
\end{equation}

For the first few values of $N$,
one finds the isomorphisms listed in Table \ref{tab:modular_groups}.

\begin{table}[ht!]
    \centering
    \renewcommand{\arraystretch}{1.3}
    \begin{tabular}{c c c}
        \toprule
        \textbf{Level} ($N$) & \textbf{Inhomogeneous Group} ($\Gamma_N$) & \textbf{Homogeneous Group} ($\Gamma'_N$) \\
        \midrule
        $2$ & $S_3$ & $S_3$ \\
        $3$ & $A_4$ & $T' \cong A'_4$ \\
        $4$ & $S_4$ & $S'_4$ \\
        $5$ & $A_5$ & $A'_5$ \\
        \bottomrule
    \end{tabular}
    \caption{Isomorphisms of the finite inhomogeneous modular groups $\Gamma_N \cong PSL(2, \mathbb{Z}_N)$ and their homogeneous double covers $\Gamma'_N \cong SL(2, \mathbb{Z}_N)$ for low levels $N \le 5$. Note that for $N=2$, the double cover is isomorphic to the group itself.}
    \label{tab:modular_groups}
\end{table}

The $SL(2,\mathbb{Z})$ modular group was already known in toroidal compactifications of string theory,
heterotic
orbifold models \cite{Ferrara:1989bc, Lerche:1989cs,Lauer:1989ax} and Type IIB superstring theories with magnetized D-branes \cite{Kobayashi:2018rad,Kobayashi:2018vbk,Ohki:2020bpo,Kikuchi:2020frp,Kikuchi:2020nxn,Kikuchi:2021ogn,Almumin:2021fbk}.

\subsection{Modular-invariant supersymmetric theories}

Considering a $\mathcal{N}=1$ supersymmetric theory invariant under the full modular group, the action in general takes the form
\begin{equation}
\mathcal{S} \;=\; \int d^4x\, d^2\theta\, d^2\bar\theta \; K(\phi_i, \bar\phi_i; \tau, \bar\tau)
\;+\;
\left(
\int d^4x\, d^2\theta \; W(\phi_i; \tau) + \text{h.c.}
\right).
\label{eq:action}
\end{equation}
Note that the modulus $\tau$ acts as the scalar component of a chiral superfield, analogous to the dilaton or moduli fields in string compactifications. Since this framework is a bottom-up supersymmetric effective field theory, rather than a full UV-complete string model, we assume the modulus $\tau$ is stabilized at its optimal phenomenological value by a suitable supergravity scalar potential in the UV completion \cite{Novichkov_2022, Knapp-Perez:2023nty, abe2025modulistabilizationfinitemodular}.

In this Section, we follow the notation of \cite{ABE2023137977}.
We know that $\tau$ parametrizes the geometry of the compact space, and a modular transformation corresponds to a reparameterization of this geometry. Since matter superfields descend from fields living on the compact space, they transform according to this reparameterization, similar to an ordinary scale symmetry:
\begin{equation}
    \phi_i \;\longrightarrow\;  (c\tau + d)^{k_i}\, \rho_i(\gamma)\, \phi_i\,,
\end{equation}
where $k_i$ is the modular weight of the superfield and $\rho_i(\gamma)$ is a unitary representation of the finite modular subgroup $\Gamma'_N$.
It is crucial to distinguish the chiral superfields from the modular forms themselves. While chiral superfields transform with the automorphy factor $(c\tau + d)$, they are not holomorphic functions of $\tau$. On the other hand, modular forms of weight $k$ and level $N$ are purely holomorphic functions $f(\tau)$ transforming as:
\begin{equation}
    f_i(\gamma\tau) = (c\tau + d)^{k}\, \rho(\gamma)_{ij}\, f_j(\tau), \qquad \gamma \in \Gamma'_N.
\end{equation}

In theories based on the standard (inhomogeneous) modular group $\Gamma_N$, the weight $k$ is restricted to be a non-negative even integer. However, by utilizing the homogeneous finite modular group $\Gamma'_N$ (the double cover), odd-weighted modular forms are also permitted.

Taking a generic superpotential, it can be expanded in powers of the matter superfields $\phi_i$ as:
\begin{equation}
    W(\tau, \phi) = \sum \left( Y_{i_1 \dots i_n}(\tau) \phi_{i_1} \dots \phi_{i_n} \right)_{\mathbf{1}}\,,
\end{equation}
where $(\dots)_{\mathbf{1}}$ denotes the contraction into a trivial singlet of $\Gamma'_N$. To ensure this modular invariance, the  $Y(\tau)$ couplings cannot be arbitrary constants; they must be modular forms of level $N$ and weight $k_Y$. To successfully compensate for the transformations of the matter fields, the modular weights and representations must satisfy the matching conditions:
\begin{equation}
    k_Y +\sum_j k_{i_j}=0 , \qquad \text{and} \qquad \rho_Y \otimes \rho_{i_1} \otimes \dots \otimes \rho_{i_n} \supset \mathbf{1}\,.
\end{equation}
Therefore, the Yukawa couplings in our theory are dynamically determined by finite-dimensional multiplets of modular forms, yielding a predictive framework for flavor physics.

Following the standard convention in the phenomenological literature of modular symmetries, we assume  minimal form for the K\"ahler potential, the kinetic terms for the chiral matter superfields $\phi_i$ are governed by

\begin{equation}
    K(\tau, \bar{\tau}; \phi_i, \bar{\phi}_i) \supset -\Lambda_K^2 \log(-i(\tau - \bar{\tau})) + \sum_i \frac{|\phi_i|^2}{(-i(\tau - \bar{\tau}))^{k_i}}\,,
\end{equation}
where $\Lambda_K$ has mass dimension one, and $k_i$ is the modular weight of the respective chiral matter superfield $\phi_i \in \{Q, u^c, d^c\}$. 

To yield canonical kinetic terms, the matter fields must be rescaled as $\phi_i \to (2 \operatorname{Im} \tau)^{k_i/2} \phi_i$. Consequently, the physical mass matrices absorb these K\"ahler metric effects. The effective Yukawa parameters are scaled accordingly:
\begin{equation}
    [Y_\phi]_{ij} \to \left(2 \operatorname{Im} \tau \right)^{-k_Y/2} [Y_\phi]_{ij}\,.
\end{equation}
Here, $k_Y$ is the weight of the modular form corresponding to that Yukawa coupling.

\section{Model}\label{sec:model}
In this Section, we present a model that successfully accommodates the observed quark mass hierarchies and mixing parameters and features purely Spontaneous CP Violation.
Following the methodology of \cite{Novichkov_2019}, we enforce a gCP that imposes all the couplings to be real. Consequently, the observed CP violation must come from a deviation of $\langle\tau\rangle$ from the CP conserving imaginary axis.

\subsection{Superfields and Superpotential}

The superfields are assigned under the $S'_4$ modular symmetry as dictated by Table \ref{tab:reps}, following the notation used in \cite{ABE2023137977}. Notice the top quark right-handed field ($t^c$) is assigned to representation $\mathbf{\hat{1}'}$, which naturally isolates its mass scale.

\begin{table}[h]
\centering
\renewcommand{\arraystretch}{1.3}
\begin{tabular}{c c c c c c c}
\hline \hline
\textbf{Field} & $Q$ & $u^c_D = (u^c, c^c)^T$ & $t^c$ & $d^c_D = (d^c, s^c)^T$ & $b^c$ & $H_u, H_d$ \\
\hline
$S'_4$ & $\mathbf{3}$ & $\mathbf{2}$ & $\mathbf{\hat{1}'}$ & $\mathbf{2}$ & $\mathbf{1'}$ & $\mathbf{1}$ \\
$k_I$ & $0$ & $-6$ & $-1$ & $-10$ & $-6$ & $0$ \\
\hline \hline
\end{tabular}
\caption{Representations and modular weights of the matter superfields.}
\label{tab:reps}
\end{table}

The invariant holomorphic superpotential is given by:
\begin{align}
\mathcal{W}_u &= \left( u^c_D \, Q \, Y_{uD}^{(6)} \right)_{\mathbf{1}} H_u + \left( t^c \, Q \, Y_{u3}^{(1)} \right)_{\mathbf{1}} H_u , \\
\mathcal{W}_d &= \left( d^c_D \, Q \, Y_{dD}^{(10)} \right)_{\mathbf{1}} H_d + \left( b^c \, Q \, Y_{d3}^{(6)} \right)_{\mathbf{1}} H_d ,
\end{align}
where $Y_{uD} = C_1 Y_{\mathbf{3'}}^{(6),1} + C_2 Y_{\mathbf{3'}}^{(6),2}$, $Y_{u3} = C_3 Y_{\mathbf{\hat{3}}}^{(1)}$, $Y_{dD} = C_3 Y_{\mathbf{3'}}^{(10),1} + C_4 Y_{\mathbf{3'}}^{(10),2} + C_5 Y_{\mathbf{3'}}^{(10),3}$, and $Y_{d3} = C_6 Y_{\mathbf{3'}}^{(6),1} + C_7 Y_{\mathbf{3'}}^{(6),2}$.

For simplicity, and to tightly constrain the parameter space, we phenomenologically restrict $C_{u3}=C_{dD,1}\equiv C_3$.

\subsection{Mass Matrices}
Using the Clebsch-Gordan coefficients of the $S'_4$ double cover, we explicitly construct the tensor contractions. For the doublet sectors ($\mathbf{2} \otimes \mathbf{3} \to \mathbf{3'}$ contracted with $\mathbf{3'}$), the invariant singlet takes the form:
\begin{equation}
\left( q^c_D \otimes Q \otimes Y \right)_{\mathbf{1}} = \frac{1}{\sqrt{3}} \left[ q^c_1 \left( \frac{\sqrt{3}}{2} Q_2 Y_2 + \frac{\sqrt{3}}{2} Q_3 Y_3 \right) + q^c_2 \left( -Q_1 Y_1 + \frac{1}{2} Q_2 Y_3 + \frac{1}{2} Q_3 Y_2 \right) \right].
\end{equation}

For the singlet sectors, $\mathbf{\hat{1}'} \otimes \mathbf{3} \to \mathbf{\hat{3}'}$ and $\mathbf{1'} \otimes \mathbf{3} \to \mathbf{3'}$ contracted with their respective modular forms yield:
\begin{equation}
\left( t^c \otimes Q \otimes Y_{u3} \right)_{\mathbf{1}} = \frac{1}{\sqrt{3}} t^c \left( Q_1 Y_{u3,1} + Q_2 Y_{u3,3} + Q_3 Y_{u3,2} \right).
\end{equation}

After canonical normalization, the rescaling of the matter fields induces diagonal factors in the physical mass matrices, controlled by the modular weights via $\mathrm{diag}((2\operatorname{Im}(\tau))^{k_i/2})$.
The exact matrices are:
\begin{equation}
M_u = \frac{v_u}{\sqrt{6}} 
\begin{pmatrix}
(2\operatorname{Im}(\tau))^{-3} & 0 & 0 \\
0 & (2\operatorname{Im}(\tau))^{-3} & 0 \\
0 & 0 & (2\operatorname{Im}(\tau))^{-1/2}
\end{pmatrix}
\begin{pmatrix}
0 & \frac{\sqrt{3}}{2} Y_{uD,2} & \frac{\sqrt{3}}{2} Y_{uD,3} \\
-Y_{uD,1} & \frac{1}{2} Y_{uD,3} & \frac{1}{2} Y_{uD,2} \\
Y_{u3,1} & Y_{u3,3} & Y_{u3,2}
\end{pmatrix},
\end{equation}

\begin{equation}
M_d = \frac{v_d}{\sqrt{6}} 
\begin{pmatrix}
(2\operatorname{Im}(\tau))^{-5} & 0 & 0 \\
0 & (2\operatorname{Im}(\tau))^{-5} & 0 \\
0 & 0 & (2\operatorname{Im}(\tau))^{-3}
\end{pmatrix}
\begin{pmatrix}
0 & \frac{\sqrt{3}}{2} Y_{dD,2} & \frac{\sqrt{3}}{2} Y_{dD,3} \\
-Y_{dD,1} & \frac{1}{2} Y_{dD,3} & \frac{1}{2} Y_{dD,2} \\
Y_{d3,1} & Y_{d3,3} & Y_{d3,2}
\end{pmatrix},
\end{equation}
where $v_u$ and $v_d$ are the Higgs vacuum expectation values.

\section{Numerical Analysis and Phenomenological Results }\label{sec:numerical-analysis}

In this Section, we perform a numerical and statistical analysis of the quark sector within the proposed $S'_4$ modular symmetry framework. The model predictions are confronted with the most recent global fit provided by the PDG \cite{ParticleDataGroup:2024cfk}, when extrapolated to the GUT scale \cite{antusch2026updatedrunningquarklepton}.
The comparison with earlier 2022 values was shown in Table \ref{tab:2024vs2022}.

To identify viable model configurations, we minimize the $\chi^2$ function defined as
\begin{equation}
\chi^2 = \sum_i \frac{\left(\mathcal{O}_i^{\text{th}} - \mathcal{O}_i^{\text{exp}}\right)^2}{\sigma_i^2}\,,
\end{equation}
where $\mathcal{O}_i^{\text{th}}$ and $\mathcal{O}_i^{\text{exp}}$ denote the theoretical predictions and experimental central values of the observables, respectively, and $\sigma_i$ their corresponding uncertainties.

We explored different model topologies by systematically varying the modular weights of the matter fields and the singlet representations assigned to the third generation. The Yukawa couplings were dynamically determined by assigning them the unique weights and representations required to maintain each term in the superpotential modular invariant and weightless.
Since our approach is strictly bottom-up, these assignments are treated as free parameters, selected purely phenomenologically to minimize the global $\chi^2$ function without presupposing a specific top-down origin.
However, we did fix the representations of our fields to have the $\mathbf{2} \oplus \mathbf{1}$ structure.

The scan was performed over parameter ranges $|C_i| \in [0.1,15]$ and $\tau$ in the fundamental domain determined via a differential evolution algorithm using precisely the Clebsch-Gordan coefficients and modular forms present in \cite{Novichkov_2021}. 

 The optimal vacuum configuration and the corresponding $\mathcal{O}(1)$ real coupling constants of this topology that minimize the $\chi^2$ function  are presented in Table \ref{tab:results}. The model has 9 free parameters (7 Constants +$\operatorname{Re}(\tau)$ + $\operatorname{Im}(\tau)$) and is consistent with the PDG 2024 values extrapolated to the GUT scale (assuming MSSM with $\tan\beta=10$ and $M_{\text{SUSY}}=3$ TeV).

Since the  ratios  $y_i/y_j$  preserve the QCD correlation during the running of RGEs, we use the error propagation of the measures at low energies.
The mixing angles and the CP phase are extremely stable under the evolution of RGE equations; therefore, they remain almost the same as in low energies.

\begin{table}[H]
\centering
\renewcommand{\arraystretch}{1.3}
\resizebox{\textwidth}{!}{
\begin{tabular}{c | c | c c c c}
\hline \hline
\textbf{Parameter} & \textbf{Value} & \textbf{Observable} & \textbf{Target (PDG 2024)} & \textbf{Model Prediction} & \textbf{Pull ($\sigma$)} \\
\hline
$\operatorname{Re}(\tau)$ & $0.00455$ & $y_u / y_c$ & $1.986 \times 10^{-3}$ & $1.981 \times 10^{-3}$ & $-0.13$ \\
$\operatorname{Im}(\tau)$ & $1.00705$ & $y_c / y_t$ & $2.810 \times 10^{-3}$ & $2.812 \times 10^{-3}$ & $+0.02$ \\
$C_1$ & $-0.3951$ & $y_d / y_s$ & $5.000 \times 10^{-2}$ & $5.009 \times 10^{-2}$ & $+0.05$ \\
$C_2$ & $+0.2181$ & $y_s / y_b$ & $1.782 \times 10^{-2}$ & $1.785 \times 10^{-2}$ & $+0.04$ \\
$C_{3}$ & $+3.7065$ & $\theta_{12} \, (\text{rad})$ & $0.2270$ & $0.2270$ & $-0.03$ \\
$C_4$ & $+1.0452$ & $\theta_{23} \, (\text{rad})$ & $0.0387$ & $0.0387$ & $-0.02$ \\
$C_5$ & $+5.5783$ & $\theta_{13} \, (\text{rad})$ & $0.00341$ & $0.00334$ & $-0.93$ \\
$C_6$ & $+1.3828$ & $\delta_{CP} \, (^\circ)$ & $65.26^\circ$ & $65.33^\circ$ & $+0.06$ \\
$C_7$ & $+0.3861$ & \multicolumn{4}{c}{\textbf{Total $\chi^2 \approx 0.891$}} \\
\hline \hline
\end{tabular}
}
\caption{The fit successfully reproduces the PDG 2024 data extrapolated to the GUT scale (assuming MSSM, $\tan\beta=10$, $M_{\text{SUSY}}=3$ TeV) with an overall $\chi^2 \approx 0.891$. All $C_i$ parameters are strictly real $\mathcal{O}(1)$ constants, indicating that CP is spontaneously broken solely by the VEV of the modulus $\tau$.}
\label{tab:results}
\end{table}

The  optimized $\langle \tau \rangle$ is in the vicinity of the symmetric fixed point $\tau_S=i$ where the full modular group is broken down to  a residual $\mathbb{Z}_4^S$ symmetry.

The quark sector involves 10 independent physical observables (masses and parameters of the Cabibbo-Kobayashi-Maskawa (CKM) mixing matrix), while the present model contains 9 free parameters. Despite the comparable number of parameters and observables, achieving a successful fit is non-trivial due to the strong experimental constraints on quark masses and mixing parameters. In particular, among the various model topologies explored, only a limited subset yields an acceptable fit, indicating that the framework retains a non-trivial degree of predictivity.

\begin{figure}[H]
    \centering
    \includegraphics[width=0.7\textwidth]{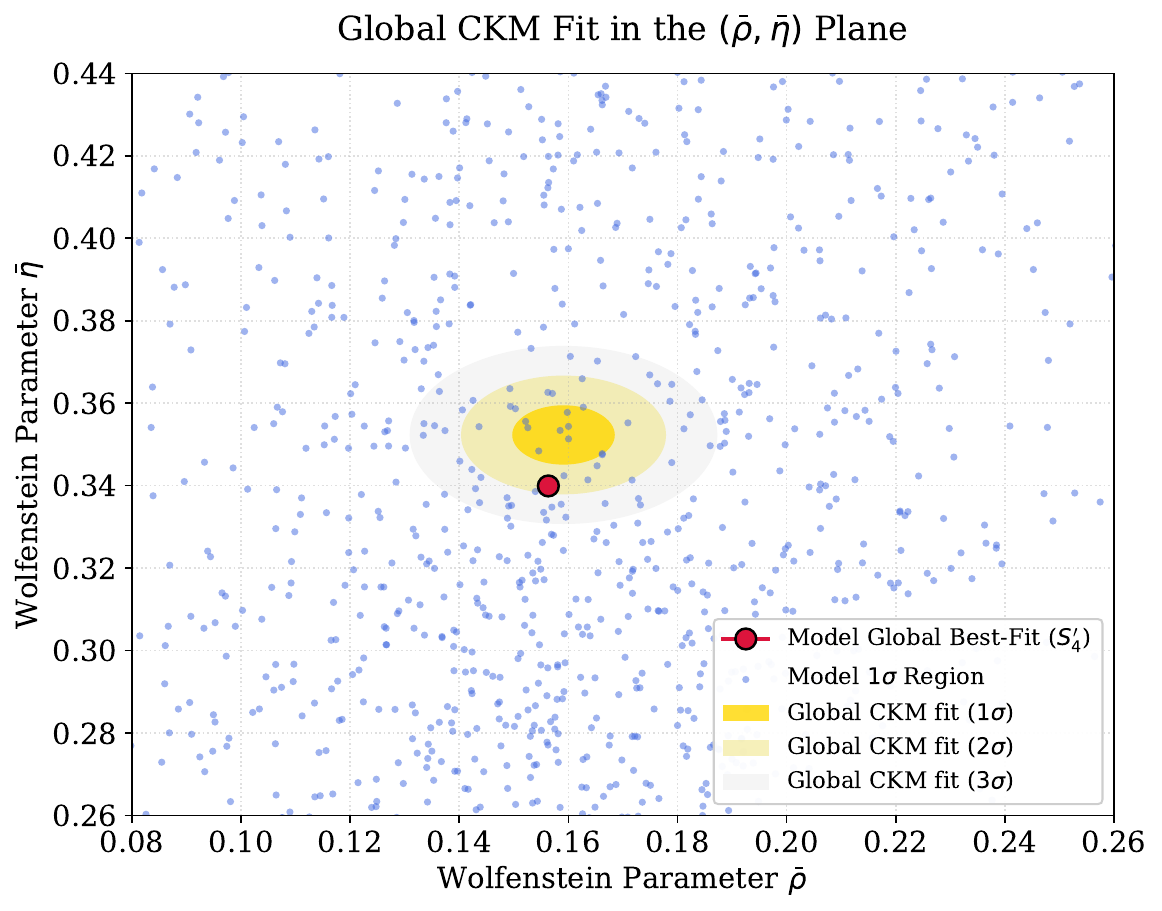}
    \caption{Projection in the $(\bar{\rho}, \bar{\eta})$ plane of the theoretical $1\sigma$ parameter space, as derived from a Gaussian sampling around the global $\chi^2$ minimum. The red marker denotes the exact global best-fit configuration of the $S'_4$ modular framework. The blue scatter points represent the theoretical $1\sigma$ region. The shaded background ellipses delineate the $1\sigma$, $2\sigma$, and $3\sigma$ experimentally allowed regions derived from the updated PDG 2024 global fit.}
    \label{fig:global_fit}
\end{figure}

To provide a rigorous analytical validation of our framework, in Fig.~\ref{fig:global_fit}, we show the $(\bar{\rho}, \bar{\eta})$ plane (parameters of the CKM in the Wolfenstein parametrization \cite{ParticleDataGroup:2024cfk} superimposed with the experimentally allowed regions, and project into the plane the model predictions according to a simultaneous Gaussian sampling of all continuous free parameters around the global $\chi^2 $ minimum.

According to the 2024 PDG global fit \cite{ParticleDataGroup:2024cfk}, the CKM parameters are determined to be 
\[
\bar{\rho}_{\text{exp}} = 0.1591 \pm 0.0094, 
\qquad 
\bar{\eta}_{\text{exp}} = 0.3523^{+0.0073}_{-0.0071}.
\]

Evaluating our model at the best fit configuration, we obtain 
\[
\bar{\rho}_{\text{model}} \simeq 0.156, 
\qquad 
\bar{\eta}_{\text{model}} \simeq 0.340,
\]
corresponding to deviations of $-0.33\sigma$ and $-1.73\sigma$, respectively.

The analysis demonstrates that the model's global minimum  resides comfortably within the $2  \sigma$ experimental boundary. The high density of valid configurations still within $1\sigma$ and $2\sigma$ regions suggests the stability of the preferred region of the $S'_4$ modular framework.  

\begin{figure}[H]
    \centering
    \includegraphics[width=0.8\textwidth]{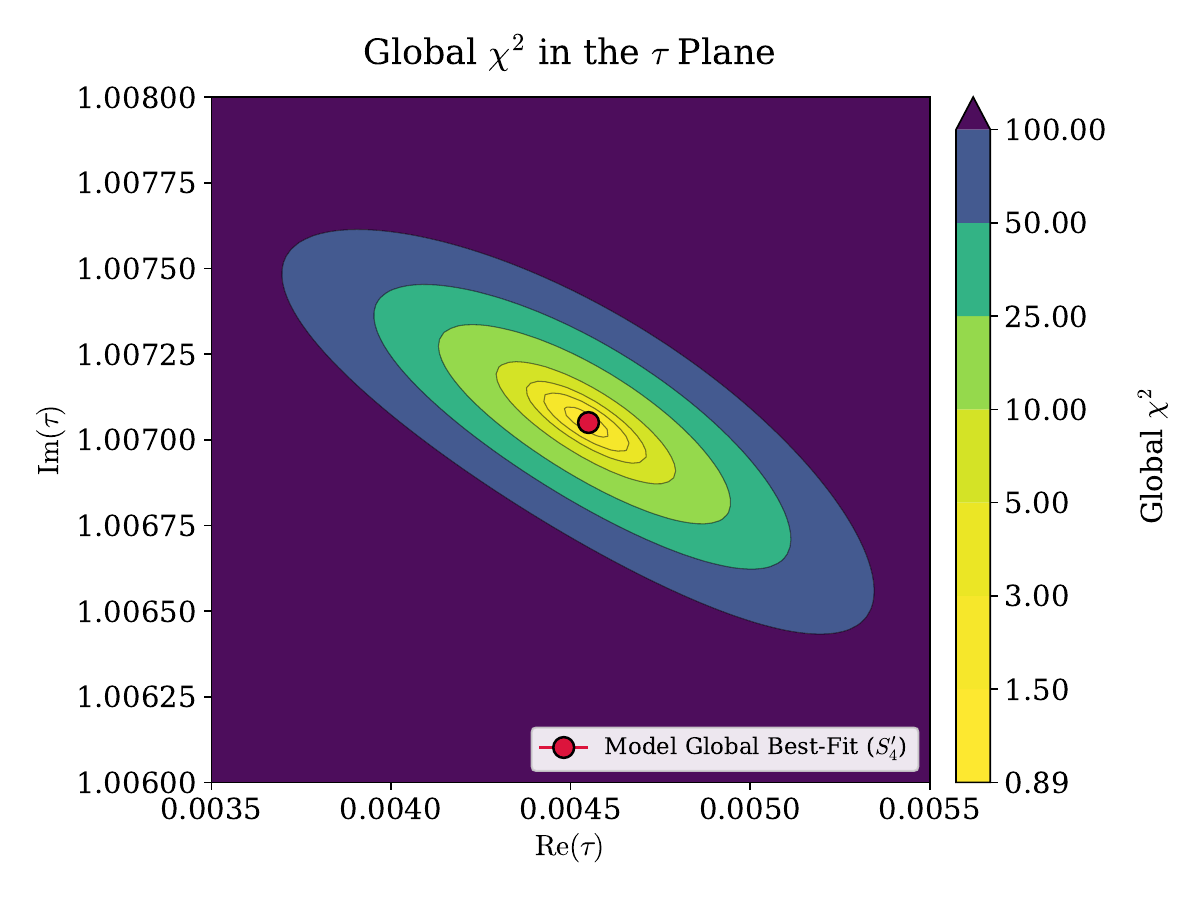}
    \caption{Topographical contour map of the global $\chi^2$ in the complex modulus plane. }
    \label{fig:vacuum_contour}
\end{figure}

As another test of the stability and the topological features of the preferred geometric vacuum, we perform a high-resolution evaluation of the global $\chi^2$ function in the immediate vicinity of the best-fit modulus, $\langle \tau \rangle$. In this analysis the real constants are kept fixed at their optimal values. The results of this analysis are presented in Fig. \ref{fig:vacuum_contour}. From this we conclude:
\begin{itemize}
    \item Our best-fit modulus resides precisely at the bottom of a well-defined local minimum with relatively steep gradients. The steep gradients surrounding the minimum indicate a high degree of predictivity of the $S'_4$ modular framework.
    \item The elliptical contours exhibit a pronounced tilt in the complex plane. This indicates a non-trivial correlation between  $\operatorname{Re} (\tau)$ and $\operatorname{Im} (\tau)$.
\end{itemize}

\begin{figure}[H]
    \centering
    \includegraphics[width=0.6\textwidth]{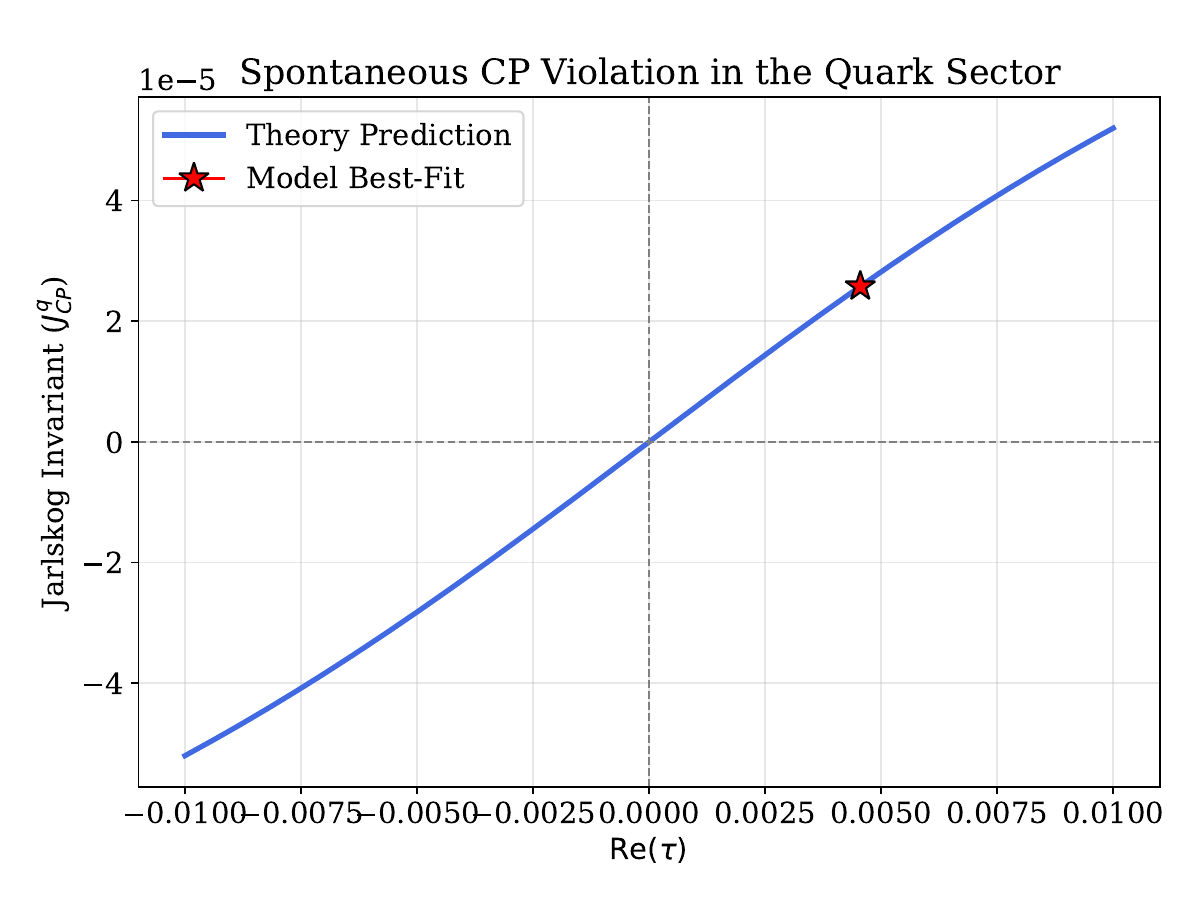}
    \caption{The Jarlskog invariant $J_{CP}^q$ as a function of $\operatorname{Re}(\tau)$. The invariant vanishes when the vacuum aligns with the imaginary axis.}
    \label{fig:scpv_proof}
\end{figure}

The optimized geometric vacuum localizes in the immediate vicinity of the symmetric fixed point $\tau_S = i$. At this fixed point, the modular symmetry is broken down to a residual $\mathbb{Z}_4^S$ subgroup and the gCP symmetry is spontaneously preserved by the vacuum. By expanding the modulus analytically around this CP-conserving point, $\tau = i + \epsilon$, we find that the CP-violating Jarlskog invariant scales linearly with the real perturbation, $J_{CP} \propto \operatorname{Re}(\epsilon)$.
Consequently, the small displacement along the real axis, $\operatorname{Re}\langle\tau\rangle \simeq 0.00455$, acts as the sole source of Spontaneous CP Violation. As shown in Fig.~\ref{fig:scpv_proof}, the theoretical prediction smoothly crosses the origin, with $J_{CP} = 0$ precisely at $\operatorname{Re}(\langle\tau\rangle) = 0$, consistent with the unbroken CP symmetry \cite{Novichkov_2019}. This links the origin of the CKM phase to a slight breaking of the residual geometry, completely decoupling the CP phase from the mass hierarchies, which are primarily related to $\operatorname{Im}(\langle\tau\rangle) \sim 1$.

\section{Conclusions}

In this work, we presented a modular flavor framework based on the double cover $S'_4$ and a General CP symmetry, to address the quark flavor problem.

All couplings were taken to be real $\mathcal{O}(1)$ parameters, so that CP violation arises spontaneously from $\langle \tau \rangle$. A key feature is the use of  the modular K\"ahler metric, together with non-trivial modular weight assignments, to generate realistic hierarchies without resorting to unjustified hierarchies.
We obtain a great fit to the PDG 2024 data extrapolated to the GUT scale (assuming MSSM, $\tan\beta=10$, $M_{\text{SUSY}}=3$ TeV) achieving $\chi^2 \simeq 0.89$.

Overall, the present framework demonstrates that modular symmetries, combined with canonical normalization effects, provide a viable and economical approach to flavor model building.

\section*{Acknowledgements}

IdMV thanks the University of Basel for hospitality.
IdMV acknowledges funding from Fundação para a Ciência e a Tecnologia (FCT) through the FCT Mobility program, and through
the project CFTP-FCT Unit 
UID/00777/2025 (\url{https://doi.org/10.54499/UID/00777/2025}).

\appendix
\section{Modular Forms and Clebsch-Gordan Coefficients of \texorpdfstring{$S'_4$}{S'4}}
For completeness, we summarize the relevant Clebsch-Gordan and tensor contractions of the $S'_4$ group and the explicit modular forms used in our framework, adopting the basis and conventions established by \cite{Novichkov_2021}.

\subsection{Tensor Products}
For the tensor products, denoting the elements of the multiplets as $\alpha_i$ and $\beta_i$, the exact contractions are given by:

\noindent For $\mathbf{1'} \otimes \mathbf{3} \to \mathbf{3'}$ and $\mathbf{\hat{1}'} \otimes \mathbf{3} \to \mathbf{\hat{3}'}$:
\begin{equation}
    \alpha_{\mathbf{1' / \hat{1}'}} \otimes \begin{pmatrix} \beta_1 \\ \beta_2 \\ \beta_3 \end{pmatrix}_{\mathbf{3}} = \alpha_1 \begin{pmatrix} \beta_1 \\ \beta_2 \\ \beta_3 \end{pmatrix}_{\mathbf{3' / \hat{3}'}}.
\end{equation}

\noindent For $\mathbf{2} \otimes \mathbf{3} \to \mathbf{3'}$:
\begin{equation}
    \begin{pmatrix} \alpha_1 \\ \alpha_2 \end{pmatrix}_{\mathbf{2}} \otimes \begin{pmatrix} \beta_1 \\ \beta_2 \\ \beta_3 \end{pmatrix}_{\mathbf{3}} \supset \begin{pmatrix} - \alpha_2\,\beta_1 \\ \frac{\sqrt{3}}{2}\alpha_1\,\beta_3 + \frac{1}{2}\alpha_2\,\beta_2 \\ \frac{\sqrt{3}}{2}\alpha_1\,\beta_2 + \frac{1}{2}\alpha_2\,\beta_3 \end{pmatrix}_{\mathbf{3'}}.
\end{equation}

\noindent For the invariant singlet contractions $\mathbf{3'} \otimes \mathbf{3'} \to \mathbf{1}$ and $\mathbf{\hat{3}'} \otimes \mathbf{\hat{3}} \to \mathbf{1}$:
\begin{equation}
    \left( \alpha \otimes \beta \right)_{\mathbf{1}} = \frac{1}{\sqrt{3}}\left(\alpha_1\beta_1 + \alpha_2\beta_3 + \alpha_3\beta_2\right).
\end{equation}

\subsection{Explicit Modular Forms}
The modular forms of level 4 can be written in terms of two ``weight 1/2'' Jacobi theta constants, $\theta(\tau)$ and $\varepsilon(\tau)$. Their $q$-expansions, in terms of $q_4 \equiv \exp(i\pi \tau/2)$, are:
\begin{align}
  \theta(\tau) &= 1 + 2 \,q_4^4 + 2 \,q_4^{16} + \dots  \\
  \varepsilon(\tau) &= 2 \,q_4^{\phantom{1}} + 2 \,q_4^9 + 2 \,q_4^{25} + \dots 
\end{align}

The specific $S'_4$ modular multiplets of weights 1, 6, and 10 used in our quark sector mass matrices are given as polynomials of these two functions:

\vspace{0.3cm}
\noindent \textbf{Weight 1:}
\begin{equation}
  Y_{\mathbf{\hat{3}}}^{(1)}(\tau) =
  \begin{pmatrix}
    \sqrt{2} \, \varepsilon \, \theta \\
    \varepsilon^2 \\
    -\theta^2
  \end{pmatrix},
\end{equation}

\vspace{0.3cm}
\noindent \textbf{Weight 6:}
\begin{align}
  Y_{\mathbf{3'},1}^{(6)}(\tau) &=
  \begin{pmatrix}
  -\frac{3}{8\sqrt{13}}\left( \theta ^{12} -3\, \varepsilon ^4\, \theta ^8 +3\, \varepsilon ^8\, \theta ^4 -\varepsilon ^{12} \right) \\
  \frac{3 \sqrt{2}}{\sqrt{13}}\left( 3 \,\varepsilon ^5\, \theta ^7 + \varepsilon ^9\, \theta ^3 \right) \\
  \frac{3 \sqrt{2}}{\sqrt{13}} \left( \varepsilon ^3\, \theta ^9 + 3\, \varepsilon ^7\, \theta ^5 \right)
  \end{pmatrix}\,, \\
  Y_{\mathbf{3'},2}^{(6)}(\tau) &=
  \begin{pmatrix}
    3\left( \varepsilon ^4\, \theta ^8 -\varepsilon ^8\, \theta ^4 \right)\\
  -\frac{3}{4\sqrt{2}}\left( \varepsilon\,  \theta ^{11} +2\, \varepsilon ^5\, \theta ^7 -3\, \varepsilon ^9\, \theta ^3 \right)\\
  \frac{3}{4\sqrt{2}}\left( 3\, \varepsilon ^3\, \theta ^9 -2\, \varepsilon ^7\, \theta ^5 -\varepsilon ^{11}\, \theta \right)
\end{pmatrix},
\end{align}

\vspace{0.3cm}
\noindent \textbf{Weight 10:}
\begin{align}
  Y_{\mathbf{3}',1}^{(10)}(\tau) &=
  \begin{pmatrix}
-\frac{3}{32\sqrt{29}} \left( \theta^{20} +59 \,\varepsilon^4 \,\theta^{16} -182 \,\varepsilon^8 \,\theta^{12} +182 \,\varepsilon^{12} \,\theta^8 -59 \,\varepsilon^{16} \,\theta^4 -\varepsilon^{20} \right) \\
3 \sqrt{\frac{2}{29}} \left( 13 \,\varepsilon^9 \,\theta^{11} +2 \,\varepsilon^{13} \,\theta^7 +\varepsilon^{17} \,\theta^3 \right) \\
3 \sqrt{\frac{2}{29}} \left( \varepsilon^3 \,\theta^{17} +2 \,\varepsilon^7 \,\theta^{13} +13 \,\varepsilon^{11} \,\theta^9 \right)
  \end{pmatrix}\,, \\
  Y_{\mathbf{3}',2}^{(10)}(\tau) &=
  \begin{pmatrix}
\frac{36}{\sqrt{13}} \left( \varepsilon^8 \,\theta^{12} -\varepsilon^{12} \,\theta^8 \right) \\
-\frac{9}{16\sqrt{26}} \left( \varepsilon \, \theta^{19} +20 \,\varepsilon^5 \,\theta^{15} +14 \,\varepsilon^9 \,\theta^{11} -28 \,\varepsilon^{13} \,\theta^7 -7 \,\varepsilon^{17} \,\theta^3 \right) \\
\frac{9}{16\sqrt{26}} \left( 7 \,\varepsilon^3 \,\theta^{17} +28 \,\varepsilon^7 \,\theta^{13} -14 \,\varepsilon^{11} \,\theta^9 -20 \,\varepsilon^{15} \,\theta^5 -\varepsilon^{19} \,\theta \right)
  \end{pmatrix}\,, \\
  Y_{\mathbf{3}',3}^{(10)}(\tau) &=
  \begin{pmatrix}
\frac{9}{8} \left( \varepsilon^4 \,\theta^{16} -3 \,\varepsilon^8 \,\theta^{12} +3 \,\varepsilon^{12} \,\theta^8 -\varepsilon^{16} \,\theta^4 \right) \\
\frac{9}{8\sqrt{2}} \left( \varepsilon^5 \,\theta^{15} -3 \,\varepsilon^9 \,\theta^{11} +3 \,\varepsilon^{13} \,\theta^7 -\varepsilon^{17} \,\theta^3 \right) \\
-\frac{9}{8\sqrt{2}} \left( \varepsilon^3 \,\theta^{17} -3 \,\varepsilon^7 \,\theta^{13} +3 \,\varepsilon^{11} \,\theta^9 -\varepsilon^{15} \,\theta^5 \right)
  \end{pmatrix}\,.
\end{align}

\printbibliography

\end{document}